\documentclass[iop,revtex4]{emulateapj}
\usepackage{lineno}

\newcommand{\kms}{km~s$^{-1}$}
\newcommand{\msun}{${\cal M}_\odot$}

\begin{document}


\title{The  Quadruple System HIP 45734}

\author{Andrei Tokovinin}
\affiliation{Cerro Tololo Inter-American Observatory,\footnote{National Science Foundation's 
 National Optical-Infrared Astronomy Research Laboratory} Casilla 603, La Serena, Chile}
\email{atokovinin@ctio.noao.edu}

\begin{abstract}
HIP 45734 is  a quadruple system of 2+2 architecture  located at 68 pc
from the  Sun.  The  outer 9\arcsec ~system A,B has a period  of $\sim
  10^4$ yr.  The subsystem Aa,Ab is  a visual binary with a period of 20.1
years  and an  eccentricity of  0.78.   Its periastron  in 2019.1  was
observed   spectroscopically,  yielding   masses   (1.10$\pm$0.04  and
0.98$\pm$0.03 \msun)  and orbital parallax,  14.90$\pm$0.37\,mas.  The
masses, luminosities, and colors approximately agree with evolutionary
models  of main  sequence stars.   The component  Aa has  a detectable
lithium  line, whereas  in  Ab it  is  absent.  The  pair  Ba,Bb is  a
single-lined spectroscopic binary with a  period of 0.55552 day and an
orbital  inclination   of  $\sim$45\degr  ~derived   by  modeling  the
rotationally broadened line profile with ``flat bottom''.  The mass of
Bb is  $\sim$0.4 \msun.   The star B  is chromospherically  active (an
x-ray  source); its  flux  is  modulated with  the  orbital period  by
starspots,  in addition  to  occasional flares.   The  system   is
  probably older than $\sim$600\,Myr; it does not belong to any known
moving group.
\end{abstract}

\maketitle

\section{Introduction}
\label{sec:intro}

Discovery of  a population of young chromospherically  active stars in
the solar neighborhood by their x-ray radiation in the 1980s and 1990s
stimulated   follow--up  observations   to  determine   their  physical
parameters,  ages, rotation,  kinematics,  etc.  Many  of these  stars
belong to young moving  groups and associations \citep{SACY}. However,
a large fraction (40\%) are old short-period binaries \citep{Makarov2003}.
Recently  renewed interest  in nearby  young stars  is stemmed  by the
search for  exoplanets using  high-contrast imagers because young
self-luminous planets are easier to detect compared to their older and
cooler counterparts \citep[e.g.,][]{SPOTS}.

The object  of this  note is a  9\arcsec ~visual binary  discovered by
J.~Hershel in 1837 and known  as HJ~4214.  Its secondary component
  B is  called a ``T Tau-type Star''  \citep{SACY}, probably because,
apart from  being chromospherically active,  the stars are  located in
the sky  close to  the Chamaeleon star  forming region.   However, the
distance, the  fast proper motion  (PM), and the radial  velocity (RV)
make it clear  that this pair is unrelated to  the molecular cloud and
only projects onto it.   Common identifications and main parameters of
the two  visual components A  and B are given  in Table~\ref{tab:par}.
The  {\it  Gaia}  astrometry  of   the  star  A  is  affected  by  its
acceleration, leading to a larger  parallax error compared to the star
B.  The $V$ magnitudes  in Table~\ref{tab:par} are calculated from the
{\it  Gaia} DR2  photometry using  the prescription  given on  its web
site.\footnote{See Chapter  5.3.7 of  {\it Gaia} DR2  documentation at
  \url{https://gea.esac.esa.int/archive/documentation/GDR2/}.   }  The
{\it  Tycho} photometry  gives 8.41  and 9.66  mag for  A and  B, both
fainter compared to {\it Gaia}.   Simbad gives for A the $V$ magnitude
of the combined light of A and B, 8.05 mag.

\begin{deluxetable}{l c c  }
\tabletypesize{\scriptsize}     

\tablecaption{Main Parameters of HIP 45734
\label{tab:par} }  
\tablewidth{0pt}                                   
\tablehead{                                                                     
\colhead{Parameter} & 
\colhead{A} & 
\colhead{B} 
}
\startdata
WDS         & \multicolumn{2}{c}{J09194$-$7739, HJ~4214, KOH 83} \\
Identifiers & HIP 45734 & TYC 9399-2452-1 \\
            & HD 81485  &   	RX J0919.4-7738      \\ 
PM (mas~yr$^{-1}$) &  $-$106.8,  68.5 &  $-$107.9,  70.8 \\
Parallax (mas)\tablenotemark{a} & 14.53 $\pm$ 0.15 & 14.66 $\pm$ 0.03 \\
Spectral type & G7V & K0IVe \\
$V$ (mag) & 8.31 & 9.44 \\
$G$ (mag) & 8.17 & 9.24 \\
$K$ (mag) & 6.78 & 7.44 \\
RV (\kms) & 4.9 & 7.7 
\enddata
\tablenotetext{b}{Proper motion and parallax are 
  from the {\it Gaia} DR2 \citep{Gaia}.}
\end{deluxetable}

\begin{figure}
\plotone{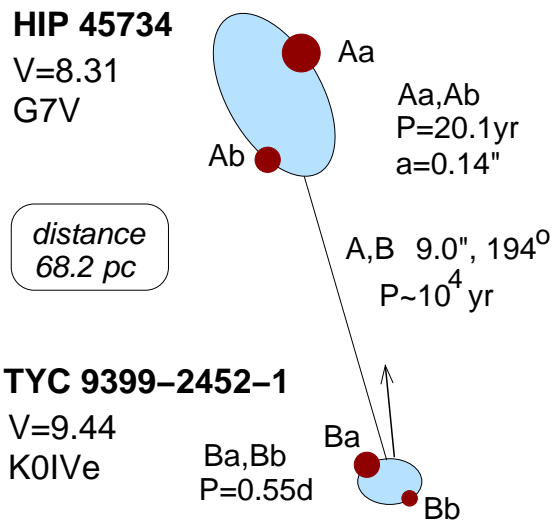}
\caption{Architecture of the HIP~45734 quadruple system (not to scale).
\label{fig:diag}
}
\end{figure}

These stars, believed to be  pre-main sequence (PMS), are featured in
the large  survey of \citet{SACY}.   The equivalent width (EW)  of the
lithium  line  in  A  and  B  was  measured  at  0.06  and  0.05  \AA,
respectively.  \citet{Covino1997}  took spectra of  HIP 45734 together
with other presumably  young stars in Chamaeleon with  a resolution of
10,000.  They  confirmed  presence of the  lithium line  and noted
that  the southern  star  B  has emissions  in  the H$\alpha$,  helium
5876~\AA ~and sodium  D lines.  Broad lines of the  star B resembled a
blend,  making Covino  et  al.   think  that B  is a  double-lined
spectroscopic  binary.    However,  no  follow-up   observations  were
conducted  to  determine the  orbit.   \citet{Desidera2006} also  took
high-resolution spectra of both visual components.  They have not seen
B as double-lined, but confirmed  the strong H$\alpha$ emission in its
spectrum and its very high chromospheric activity index; the H$\alpha$
emission in A  was ``much lower''.  Thinking that  B is a double-lined
binary, \citet{SPOTS}  looked for low-mass companions  around it using
high-contrast  adaptive  optics, but  found  only  a faint  background
source at 4\farcs3 separation.

The  star   A  was  resolved  by  \citet{Koehler2001}   into  a  close
binary. Its first  visual orbit with a period of  19.8 yr was computed
by  \citet{Tokovinin2015}.  My  interest in  this quadruple  system is
twofold.  First, it  belongs to the sample of  solar-type stars within
67\,pc \citep{FG67}, although the {\it Gaia} parallax now puts it just
outside its distance limit.  Spectroscopic observations were conducted
to  determine   the  unknown   period  of  Ba,Bb,   complementing  the
multiplicity statistics in this sample.   On the other hand, the orbit
of the  interferometric pair Aa,Ab  offered the prospect  of measuring
masses  of young,  possibly PMS,  stars to  test  stellar evolutionary
models. Both goals  are now reached  and make  the subject of this
  paper.

Figure~\ref{fig:diag} illustrates  the architecture of  this quadruple
system according  to this  study.  The pair  Aa,Ab is a  visual binary
with  an  eccentric 20.1  yr  orbit.  It  is  composed  of normal  and
apparently  inactive dwarfs  with masses  of 1.12  and 1.0  \msun.  In
contrast, Ba,Bb is a close  spectroscopic binary with a period of only
0.55  day. The  star  Ba, similar  in mass  to  Ab, has  a fast  axial
rotation synchronized  with the orbit.   Naturally, it is  also highly
active.   The mass  of the  spectroscopic  secondary Bb  is about  0.4
\msun.  Despite the  short period, this pair is not  in contact; it is
not eclipsing owing to the large inclination.

Observational   data    and   methods   are    recalled   briefly   in
Section~\ref{sec:obs}.   Then in Sections~\ref{sec:A}  and \ref{sec:B}
the  orbits of both  inner subsystems  are given;  the outer  orbit is
discussed  in  Section~\ref{sec:AB}.  Section~\ref{sec:model}  matches
stellar parameter  to the evolutionary models.  A  short discussion in
Section~\ref{sec:sum} closes the paper.

\section{Observational Data}
\label{sec:obs}

\begin{deluxetable}{l c c c   }
\tabletypesize{\scriptsize}     
\tablecaption{Parameters of the CCF profiles
\label{tab:EW} }  
\tablewidth{0pt}                                   
\tablehead{                                                                     
\colhead{Parameter} & 
\colhead{Aa} &
\colhead{Ab} & 
\colhead{B} 
}
\startdata
Amplitude                   & 0.242 & 0.161 & 0.053 \\
$\sigma$  (km~s$^{-1}$)     & 4.28   & 4.07 & 38.7 \\
Amplitude$\times \sigma$  (km~s$^{-1}$)    & 1.03 & 0.66 & 2.05 \\
$V \sin i$  (km~s$^{-1}$) &  4.7   & 4.0   &  67.4
\enddata
\end{deluxetable}

High-resolution  ($R\sim 80,000)$ spectra  of the  components A  and B
 (10  and 15,  respectively) were taken  with the  CHIRON optical
echelle  spectrometer  \citep{CHIRON} in  2015  \citep{survey} and  in
2018--2019, in the service mode.   They cover the wavelength range
  from  415\,nm  to  880\,nm  in  53  orders.   The  spectrograph  is
fiber-fed  by  the 1.5  m  telescope  operated  by Small  \&  Moderate
Aperture  Research  Telescope  System  (SMARTS)  Consortium.\footnote{
  \url{http://www.astro.yale.edu/smarts/}}  The data  analysis closely
follows  previous work  \citep{paper1}.  A  cross-correlation function
(CCF) of the  reduced spectrum with a binary mask   using lines in
  the range from  450\,nm to 650\,nm allows us to  measure the RV and
to estimate  other parameters such as  the line width  and the related
projected  axial  rotation  $V  \sin  i$. Average  parameters  of  the
Gaussian   functions  that   approximate  the   CCFs  are   listed  in
Table~\ref{tab:EW}.   The rotational velocities  were deduced from
  the  width  of  the  CCF  dips  for Aa  and  Ab  and  determined  in
  Section~\ref{sec:ccf} for B.  The  RVs and their residuals from the
orbits are given below.

The  CHIRON spectra  show no  emissions in  the Bahlmer  lines  of the
component  A, while  these lines  in the  component B  are practically
absent, being completely filled  by emission.  No other emission lines
are detectable in either component.

The positional measurements of the pair Aa,Ab used for the calculation
of its combined  orbit are, mostly, made by the  speckle camera at the
4.1   m  Southern  Astrophysical   Research  (SOAR)   telescope.   The
instrument  and data  processing are  described by  \citet{HRCAM}. The
latest series of measurements and references to prior publications can
be  found  in  \citet{SAM19}.   The  Washington  Double  Star  Catalog
\citep[WDS,][]{WDS}  was consulted for  the published  measurements of
the outer  and inner  resolved pairs.    Since the first  orbit of
  Aa,Ab was computed in 2014, 10 new speckle measurements covering the
  periastron  have been  made.  

As in the  previous papers of this series,  orbital elements and their
errors  were  determined  by   the  least-squares  fits  with  weights
inversely proportional to the square  of adopted errors.  The IDL code
{\tt                                          orbit}\footnote{Codebase:
  \url{http://www.ctio.noao.edu/\~{}atokovin/orbit/}}     was     used
\citep{orbit}.

\section{Subsystem A{\rm a},A{\rm b}}
\label{sec:A}

\begin{figure*}
\plotone{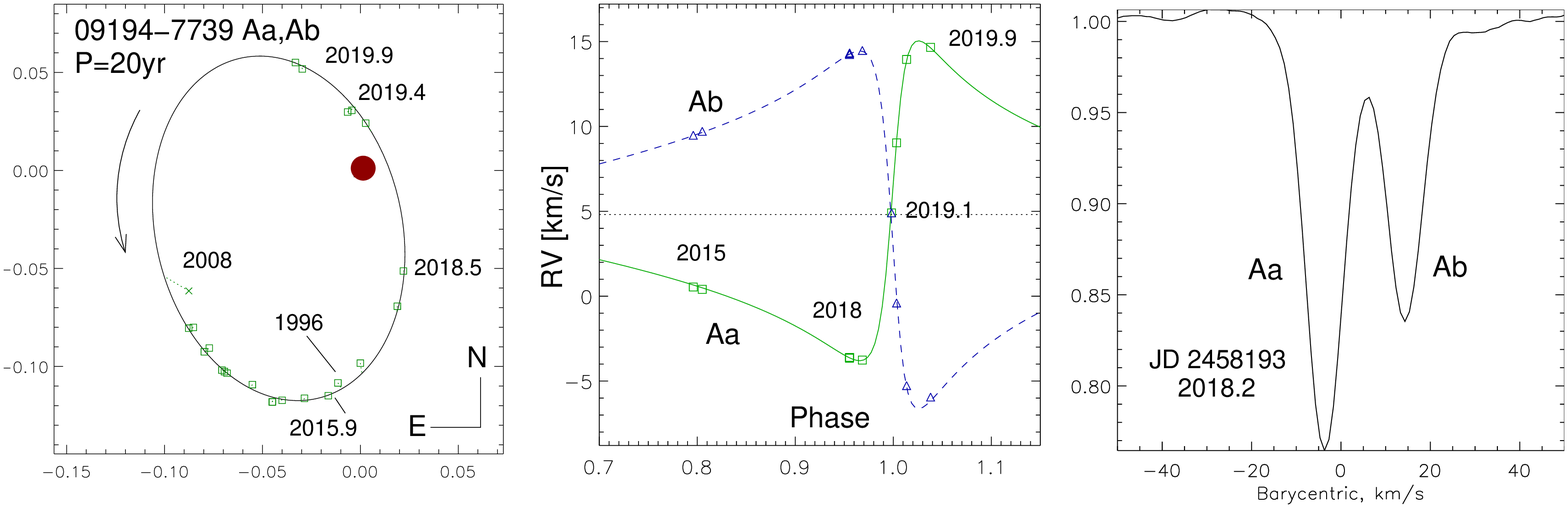}
\caption{Visual and  spectroscopic orbit of HIP  45734 Aa,Ab (KOH~83).
  Left: orbit in the sky (scale in arsceconds, Aa is at
  the coordinate origin), center: fragment of the RV curve, right: CCF
  in 2018.2 with two resolved dips.
\label{fig:A}
}
\end{figure*}


\begin{deluxetable}{l c  }
\tabletypesize{\scriptsize}     
\tablecaption{Combined Orbit of A{\rm a},A{\rm b}
\label{tab:vb} }  
\tablewidth{0pt}                                   
\tablehead{                                                                     
\colhead{Parameter} & 
\colhead{Value} 
}
\startdata
Period $P$ (yr) &  20.10 $\pm$  0.29 \\
Periastron $T_0$ (yr) & 2019.103  $\pm$          0.004 \\
Eccentricity $e$ & 0.7818  $\pm$        0.0038 \\
Semimajor axis $a$ (arcsec) &  0.1405  $\pm$         0.0019 \\
Node $\Omega$ (deg) &   21.6    $\pm$        0.7 \\
Longitude $\omega$ (deg) &   276.3  $\pm$   0.3 \\
Inclination $i$ (degr) &  63.0  $\pm$          0.6 \\
Primary amplitude $K_1$ (km~s$^{-1}$) &  9.420    $\pm$  0.042 \\
Secondary amplitude $K_2$ (km~s$^{-1}$) &  10.537    $\pm$  0.073 \\
$\gamma$ velocity  (km~s$^{-1}$) &  4.813     $\pm$  0.029 \\
R.M.S. residuals  (km~s$^{-1}$) &  0.05,     0.17 \\
$M_{\rm Aa}$ (${\cal M}_\odot$) & 1.12$\pm$0.04 \\ 
$M_{\rm Ab}$ (${\cal M}_\odot$) & 1.00$\pm$0.03 
\enddata
\end{deluxetable}

\begin{deluxetable}{l rrr rr c}    
\tabletypesize{\scriptsize}     
\tablecaption{Measurements and residuals of  A{\rm a},A{\rm b}
\label{tab:speckle}          }
\tablewidth{0pt}                                   
\tablehead{                                                                     
\colhead{Date} & 
\colhead{$\theta$} & 
\colhead{$\rho$} & 
\colhead{$\sigma_\rho$} & 
\multicolumn{2}{c}{O$-$C$_{\theta, \rho}$} &
\colhead{Ref.\tablenotemark{a}} \\
\colhead{(yr)} &
\colhead{(\degr)} &
\colhead{(\arcsec)} &
\colhead{(\arcsec)} &
\colhead{(\degr)} &
\colhead{(\arcsec)} &
}
\startdata
  1996.2420 & 173.9 &  0.1090  &  0.010 &  $-$1.2 &  $-$0.003 &K\\
  2008.1329 & 125.2 &  0.1070  &  0.050 &   6.7 &  $-$0.007 &V\\
  2010.0841 & 133.3 &  0.1170  &  0.010 &   0.5 &  $-$0.002 &T\\
  2010.0841 & 132.7 &  0.1190  &  0.010 &  $-$0.2 &  $-$0.000 &T\\
  2011.0353 & 139.4 &  0.1221  &  0.002 &  $-$0.0 &   0.000 &S\\
  2011.0353 & 139.7 &  0.1190  &  0.002 &   0.3 &  $-$0.003 &S\\
  2012.1003 & 146.8 &  0.1238  &  0.002 &   0.3 &   0.000 &S\\
  2012.1003 & 145.4 &  0.1240  &  0.002 &  $-$1.1 &   0.000 &S\\
  2012.1003 & 146.1 &  0.1239  &  0.002 &  $-$0.4 &   0.000 &S\\
  2013.1270 & 153.4 &  0.1224  &  0.005 &   0.2 &  $-$0.002 &S\\
  2014.0418 & 159.3 &  0.1264  &  0.002 &   0.2 &   0.002 &S\\
  2014.0418 & 159.4 &  0.1263  &  0.002 &   0.2 &   0.002 &S\\
  2014.3010 & 161.4 &  0.1239  &  0.002 &   0.5 &   0.000 &S\\
  2015.1036 & 166.4 &  0.1197  &  0.002 &   0.1 &  $-$0.001 &S\\
  2015.9117 & 172.1 &  0.1161  &  0.002 &   0.0 &   0.000 &S\\
  2016.9577 & 180.2 &  0.0983  &  0.009 &  $-$0.4 &  $-$0.005 &S\\
  2018.0850 & 195.5 &  0.0718  &  0.005 &   0.9 &  $-$0.003 &S\\
  2018.4815 & 203.4 &  0.0558  &  0.002 &  $-$0.4 &  0.000 &S\\
  2019.2099 & 353.9 &  0.0243  &  0.009 &   9.6 &   0.004 &S\\
  2019.3717 &  12.6 &  0.0305  &  0.009 &   4.9 &  $-$0.003 &S\\
  2019.3717 &   8.4 &  0.0310  &  0.009 &   0.7 &  $-$0.003 &S\\
  2019.8573 &  30.1 &  0.0597  &  0.002 &   0.6 &  $-$0.002 &S \\
  2019.9530 &  31.3 &  0.0643  &  0.002 & $-$0.6 & $-$0.001 & S  
\enddata 
\tablenotetext{a}{
K: \citet{Koehler2001}; 
V: \citet{Vogt2012}; 
T: \citet{Tokovinin2010};
S: speckle interferometry at SOAR;
}
\end{deluxetable}

\begin{deluxetable}{c rr rr}    
\tabletypesize{\scriptsize}     
\tablecaption{Radial Velocities of A{\rm a},A{\rm b}
\label{tab:rvA}          }
\tablewidth{0pt}                                   
\tablehead{                                                                     
\colhead{Date} & 
\colhead{$V_1$} & 
\colhead{(O$-$C)$_1$} & 
\colhead{$V_2$} & 
\colhead{(O$-$C)$_2$ } \\
\colhead{(JD $+$2400000)} &
\multicolumn{2}{c}{(km s$^{-1}$)}  &
\multicolumn{2}{c}{(km s$^{-1}$)}  
}
\startdata
 57026.6730 &    0.54 &    $-$0.11 &    9.48 &     0.02 \\ 
 57093.7885 &    0.40 &    $-$0.09 &    9.71 &     0.06 \\  
 58193.5840 & $-$3.62 &       0.01 &   14.30 &     0.04 \\  
 58194.5826 & $-$3.65 &    $-$0.03 &   14.24 &  $-$0.02 \\  
 58195.6018 & $-$3.60 &       0.03 &   14.34 &    0.08 \\  
 58290.5288 & $-$3.76 &       0.02 &   14.48 &     0.05 \\  
 58508.7680 &    4.91 &    $-$0.35 &    4.91 &     0.59 \\  
 58546.6049 &    9.04 &    $-$0.10 &   $-$0.40 &    $-$0.38 \\  
 58621.5882 &   13.95 &       0.04 &   $-$5.27 &     0.09 \\  
 58800.8457 &   14.66 &       0.03 &   $-$5.94 &     0.23  
\enddata 
\end{deluxetable}

The pair Aa,Ab  was first resolved by \citet{Koehler2001}  in 1996. It
was  not  measured in  the  following  years,  missing the  periastron
passage  in 1999.  The  next measurements  were made  only in  2008 by
\citet{Vogt2012} and in 2010  by \citet{Tokovinin2010}.  This pair has
been followed by speckle interferometry  at SOAR since 2011. The first
visual  orbit of  Aa,Ab with  a  period of  19.8 yr  was published  by
\citet{Tokovinin2015};  it is  refined  here using  both   10  new
  position  measurements   and  10  RVs   (Table~\ref{tab:vb}).   The
position   measurements    and   their   residuals    are   given   in
Table~\ref{tab:speckle}.  All  speckle measurements made  at SOAR were
re-examined and  adjusted for slightly  revised calibration parameters
of  each  observing run,  derived  from  wide  pairs as  described  in
\citet{HRCAM}.  The errors are assigned  based on the data quality and
used  for setting  weights  inversely proportional  to  the square  of
errors.  The weighted rms residuals from the new orbit are 1.4\,mas in
both  directions.  The  orbital period  is mostly  constrained  by the
first  observation in  1996 and  the  observations of  the same  orbit
segment one revolution  later, in 2015; the updated  period is 20.1 yr
(Figure~\ref{fig:A}).

The spectrum of  A taken with CHIRON in 2015  had blended narrow lines
of both  components. In 2018  the lines separated further  apart, then
closed again and swapped after  passing the periastron in 2019.1.  The
RVs of  both components determined  from the CHIRON spectra  and their
residuals to the orbit are given in Table~\ref{tab:rvA}.   The rms
  residuals are  0.05 and 0.17 \kms  ~for Aa and  Ab, respectively; we
  set the  RV errors  to 0.07  and 0.15 \kms  ~to balance  the relative
  weights  of the  RV and  speckle data  in the  combined  orbit.  RVs
  derived from the blended  CCF dips in 2015 were given  a lower weight by
  assigning the errors of 0.2 \kms.

The accurate  {\it Gaia}  DR2 parallax of  the star  B, 14.66$\pm$0.03
mas,  and the  orbital elements  yield the  mass sum  of 2.18$\pm$0.09
\msun. The mass sum error is mostly produced by the uncertainty of the
$a^3/P^2$ ratio, estimated here by  fitting 100 orbits where the input
data  are randomly  perturbed  by their  nominal  errors. This  method
accounts for the correlations  between all elements.  For example, the
inclination  $i$ and the  semimajor axis  $a$ are  strongly correlated
owing to  the unfavorable  orbit orientation. On  the other  hand, the
{\it  Gaia} parallax error  does not  contribute substantially  to the
overall  mass error.  The  spectroscopic mass  ratio $q_{\rm  Aa,Ab} =
0.894\pm0.007$ leads  to the  individual masses of  $1.15 \pm  0.05$ and
$1.03 \pm 0.04$ for Aa and Ab, respectively. 

Independently   of   the    trigonometric   parallax,   the   combined
spectro-interferometrc orbit leads to  the masses of 1.10$\pm$0.04 and
0.98$\pm$0.03  \msun ~for  Aa and  Ab, while  the orbital  parallax is
14.90$\pm$0.37  mas.   I adopt  the  masses  of  1.12 and  1.0  \msun,
compatible  with  both estimates  within  their  errors.  The  orbital
parallax agrees within  its error with the {\it  Gaia} parallax of the
component  B.  The  mass measurement  from the  visual orbit  and {\it
  Gaia} parallax  is slightly less accurate than  the mass measurement
from the combined orbit.  If Aa,Ab were a simple binary, rather than a
quadruple, the situation would  be worse because {\it Gaia} parallaxes
of unresolved close binaries are less accurate and often biased.  This
will  be  corrected  in  the   future  {\it  Gaia}  data  releases  by
incorporating orbital motion into  the astrometric model and using 
ground-based measurements  to extend the orbit coverage.

I compared  the PM anomaly  of the star  A (difference between  its PM
measured by {\it Gaia} with the mean PM derived from the Hipparcos and
{\it  Gaia}  positions)  given  by \citet{Brandt2018},  $(1.4,  -0.8)$
mas~yr$^{-1}$, with its  value calculated from the orbit  of Aa,Ab and
scaled  by   the wobble  factor  $f=-0.10$  (see   below),  $(1.1,  -0.3)$
mas~yr$^{-1}$.  The agreement  is satisfactory, considering that
the {\it Gaia} PMs can be biased by the orbital acceleration of A.

\section{Subsystem B{\rm a},B{\rm b}}
\label{sec:B}

\subsection{Spectroscopic Orbit}

\begin{figure}
\plotone{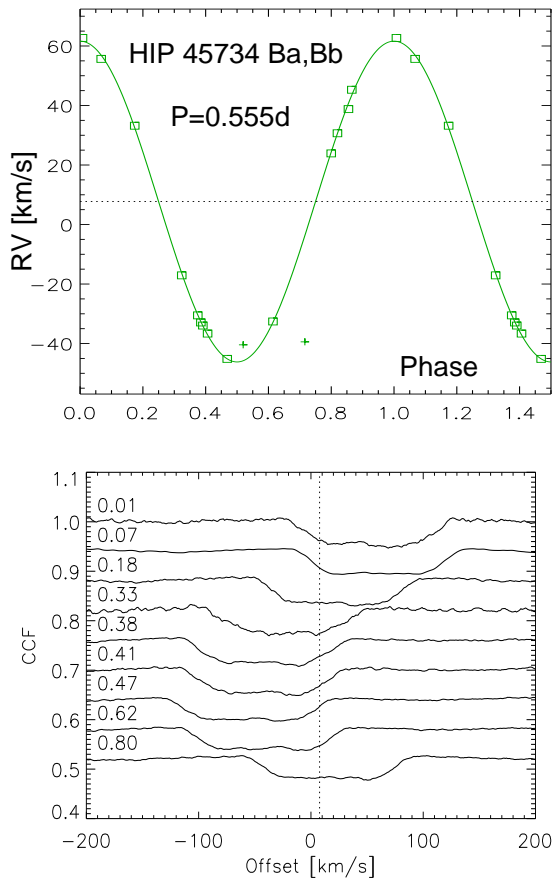}
\caption{Radial velocity curve of the  subsystem Ba,Bb (top).  The lower
  plot shows  CCF profiles  ordered by the  orbital phase,  written near
  each curve; the vertical dotted line marks the systemic velocity.
\label{fig:B}
}
\end{figure}

\begin{deluxetable}{l c  }
\tabletypesize{\scriptsize}     
\tablecaption{Spectroscopic Orbit of B{\rm a},B{\rm b}
\label{tab:sb} }  
\tablewidth{0pt}                                   
\tablehead{                                                                     
\colhead{Parameter} & 
\colhead{Value} 
}
\startdata
Period $P$ (day) &  0.5555195 $\pm$  0.000007 \\
Periastron $T_0$ (JD) & 2458195.6005  $\pm$          0.0007 \\
Eccentricity $e$ & 0 (fixed) \\
Longitude $\omega$ (deg) & 0 (fixed) \\
Primary ampl. $K_1$ (km~s$^{-1}$) &  53.94   $\pm$  0.43 \\
$\gamma$ velocity  (km~s$^{-1}$) &  7.70    $\pm$  0.28 \\
R.M.S. residuals  (km~s$^{-1}$) &  0.90 
\enddata
\end{deluxetable}

\begin{deluxetable}{c rr }    
\tabletypesize{\scriptsize}     
\tablecaption{Radial Velocities of B{\rm a}
\label{tab:rvB}          }
\tablewidth{0pt}                                   
\tablehead{                                                                     
\colhead{Date} & 
\colhead{RV} & 
\colhead{(O$-$C) } \\
\colhead{(JD $+$2400000)} &
\multicolumn{2}{c}{(km s$^{-1}$)}  
}
\startdata
 57026.6764 &   23.90 &     $-$0.46  \\
 57093.7912 &  $-$32.58 &      0.26  \\
 57098.7099 &  $-$45.15 &     $-$0.01  \\
 58193.5867 &  $-$30.52 &     $-$0.13  \\
 58194.5861 &   33.21 &      0.73  \\
 58195.6047 &   62.70 &      1.14  \\
 58228.5560 &  $-$17.09 &     $-$0.69  \\
 58232.4904 &  $-$36.65 &      0.44  \\
 58284.5209 &   55.66 &     $-$1.23  \\
 58287.4749 &  $-$32.92 &     $-$0.26  \\
 58287.4784 &  $-$33.95 &      0.09  \\
 58290.5196 &   45.29 &      1.78  \\
 58546.6081 &   38.79 &     $-$1.92  \\
 58621.5838 &   30.67 &      0.05  \\
 58800.8497 &  $-$40.43 &      5.36  
\enddata 
\end{deluxetable}

The   component   B  was   considered   a   spectroscopic  binary   by
\citet{Covino1997} who  believed to have seen double  lines.  My first
observation with  CHIRON in 2015  \citep{survey} have shown  a strange
broad CCF dip  with a flat bottom that resembled  blended lines of two
stars.   However, further spectroscopic  monitoring revealed  that the
CCF profile has  a variable RV without changing  its shape. Therefore,
the star B is a  single-lined spectroscopic binary.  Some spectra of B
were taken  with a  resolution of 30,000,  sufficient for  measuring the
RVs.

The CHIRON RVs lead to  the unique orbital solution with $P=0.555$ day
(Figure~\ref{fig:B}  and  Table~\ref{tab:sb}).   Approximation of  the
$\Pi$-shaped CCF by a Gaussian curve is poor, so the dip parameters in
Table~\ref{tab:EW} derived  from such approximation  are inaccurate. I
determined  the  average CCF  profile  and tried  to  fit  it to  each
individual CCF. However, the  resulting RVs were practically identical
to the  RVs derived by the Gaussian  fits, hence I use  the latter for
consistency.   The  RVs  and  residuals  to the  orbit  are  given  in
Table~\ref{tab:rvB}.   Despite  the  shallow  and wide  CCF,  the  rms
residuals to the orbit are moderate,  only 0.9 \kms;  RV errors of
  1 \kms  ~are assumed.  This  provides an indirect evidence  that the
secondary  component Bb  does not  contribute to  the CCF.   The lower
panel of Figure~\ref{fig:B}  shows CCFs of the component  B ordered by
the orbital phase.   The CCF shape does  not correlate with  the phase and
remains approximately constant.

Two  crosses in  Figure~\ref{fig:B} denote  the  RV of  B measured  by
\citet{Desidera2006}  and the  last CHIRON  RV measured  in  2019.  If
these  RVs  are  used  to   fit  the  orbit,  the  residuals  increase
substantially.   A  small  drift  of  the period  would  explain  this
inconsistency,  although  its relative  value  is  constrained by  the
available CHIRON RVs to within $\sim$1 ppm.

\begin{figure}
\plotone{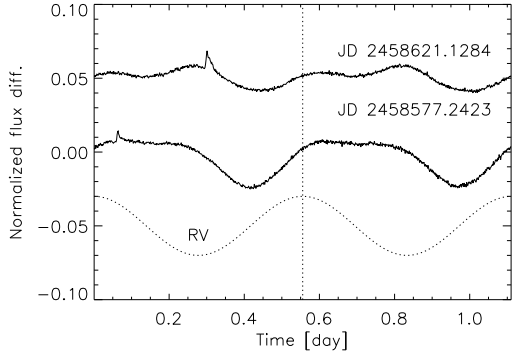}
\caption{Two  segments of TESS  light curves  with 2  mininute cadence
  phased with the  orbital period of Ba,Bb (0.55552  d).  Julian dates
  of  the segments'  start  are indicated.   The  plots are  displaced
  vertically by  0.05 to  avoid overlap.  The  dotted cosine  curve is
  proportional to  the RV variation  predicted by the orbit,  with the
  phase zero corresponding to the RV maximum.
\label{fig:lightcurve}
}
\end{figure}

\subsection{Variability}

Micro-variability  with  the period  of  0.5551  day  was detected  by
\citet{Kiraga2012}  and  attributed tentatively  to  the component  B.
Therefore,  Ba  rotates  synchronously  with the  orbit.   The  quoted
amplitude  of the  variability, 11  mmag in  $V$ and  27 mmag  in $I$,
refers to the combined light of  A and B, hence the actual variability
of B  is $\sim$2.5$\times$  larger.  It is  presumably caused  by star
spots.

The   Transiting  Exoplanet  Survey   Satellite  \citep[TESS,][]{TESS}
recently furnished accurate combined photometry  of the stars A and B.
I downloaded  the aperture  photometry of the  short-cadence sequence
provided                  by                  the                 TESS
pipeline.\footnote{https://mast.stsci.edu/portal/Mashup/Clients/Mast/Portal.html}
Flux  variation   with  the  orbital   period  of  Ba,Bb   is  obvious
(Figure~\ref{fig:lightcurve}). Although the photometric period closely
matches the orbital period, confirming the synchronous rotation of Ba,
the  shape of  the light  curve  is not  constant, reflecting  varying
distribution  of spots.   The variable  peak to  peak  flux modulation
corrected for the dilution by the light of A reaches almost $\sim$0.1.
Spikes in the  light curves show flares in  the active chromosphere of
Ba.

Migrating spots  affect the  line profile of  the fast rotator  Ba and
contribute to the  RV scatter.  Note that the CCF  of the component Ba
(Figure~\ref{fig:B})  is  slightly  asymmetric,  being  lower  on  the
right-hand side.   This asymmetry does  not change with  orbital phase
and could be of  instrumental origin, given its sub-percent amplitude.
An  asymmetry should  cause a  systematic  positive shift  of the  RV,
biasing the measured center-of-mass velocity of Ba,Bb.

\subsection{CCF Profile and Inclination}
\label{sec:ccf}

\begin{figure}
\plotone{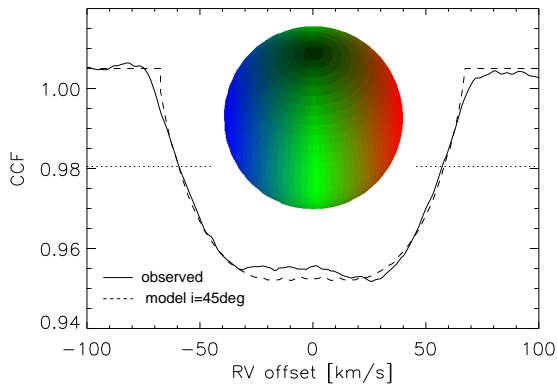}
\caption{Average CCF  of the component  B (full line) and  the modeled
  profile of a  rotating star (dash) with $i =  45\degr$ and $\alpha =
  10$. The  image illustrates the model  of a rotating  star where the
  color  reflects the  RV and  the  intensity depicts  the polar  spot
  without accounting  for the limb  darkening. The actual model  has a
  finer grid and accounts for all effects. 
\label{fig:CCF}
}
\end{figure}

The rotation of Ba is  synchronized with the orbit.  The RV amplitude,
CCF profile, and photometry, considered jointly, lead to the estimation
of the orbital  inclination $i_{\rm Ba,Bb} = 45\degr$  and the mass of
the  secondary  component,   $\sim$0.4  \msun.   First,  the  absolute
magnitude  $M_V =  5.30$  mag  and the  effective  temperature $T_e  =
5250$\,K of the star B (see Section~\ref{sec:model}) allow calculation
of  its radius  $R_*$ as  $\log  R_* =  0.5 \log(L/L_\odot)  - 2  \log
(T_e/5777)$.    With  the   bolometric  correction   of   $-0.24$  mag
appropriate for the spectral type  of B, this  formula gives $R_*  = 1.05
R_\odot$. The Ba mass of  1 \msun ~is assumed because the luminosities
of B and Ab are equal.

Chromospherically  active   binaries  of   RS  CVn  type   often  have
``flat-bottom''    line    profiles    when   the    inclination    is
small. \citet{Hatzes1996} explained this  fact by the presence of dark
polar  spots.  To investigate  this issue,  I computed  the broadening
function of a rotating inclined star by dividing its surface into many
small zones and  summing up  contributions of  visible zones to the
line profile.   A plausible  quadratic limb darkening  (intensity drop
from 1 at the center to 0.32 at the limb) was assumed.  To account for
possible  polar  spots,  my  model  includes  the  latitude  intensity
dependence as $1  + \alpha \cos \phi$ ($\phi$  is the latitude), where
$\alpha  > 0$  means dark  polar spots  and $\alpha  < 0$  means polar
brightening.  The dashed line  in Figure~\ref{fig:CCF} is computed for
$i=45\degr$ and $\alpha = 10$,  i.e.  11$\times$ dimmer at the pole
than at the equator. The exact  value of $\alpha$ does not matter when
it is  large, but  small or  zero values do  not reproduce  the ``flat
bottom'' line  profile. The model qualitatively agrees  with the shape
of the  CCF profile.  However, a  small maximum at  the center appears
only  at   inclinations  $i  \sim  20\degr$,  which   would  imply  an
unrealistically fast equatorial speed.

The Full Width at Half Maximum  (FWHM) of the CCF profile is 116 \kms.
The  FWHM of  the modeled  CCF profile  equals $2  \times 0.86  V \sin
i$.  The proportionality  coefficient is almost independent of the
  inclination:  0.86 for $i  = 35\degr$  and $i  = 45\degr$,  0.84 for
  $i=55\degr$.  Hence,   $V \sin i = 58/0.86 = 67.4$ \kms.   On the other hand,
the equatorial  speed of a star with  $R_* = 1.05 R_\odot$   and a
  period of  0.55 d is 94.7  \kms, therefore $i =  45\degr$.  At this
inclination, the orbit gives the secondary mass of $M_{\rm Bb} = 0.36$
\msun  ~if $M_{\rm  Ba} =  1$ \msun  ~is adopted.   The mass  ratio is
$q_{\rm Ba,Bb} \approx 0.4$.  Large inclination implies the absence of
eclipses, as observed.

Another estimate of the mass ratio, independent of orbital inclination
and  almost independent  of the  assumed mass  of Ba,  is  obtained by
comparison of the orbital and  rotation speeds.  For a circular orbit,
the RV amplitude of the main star Ba  is $K_1 = a \omega q \sin i/(1 +
q)$,  where $\omega =  2 \pi/P$  is the  angular speed  and $a  = 3.15
R_\odot$ is the  orbital radius, computed from the  third Kepler's law
for the  mass sum of 1.4  \msun.  On the other  hand, $V \sin  i = R_*
\omega \sin i$.  The ratio of  these equations gives $q/(1 + q) = (K_1
/V  \sin i)  \times (R_*/  a)$ or  $q =  0.36$ after  substituting the
measured and estimated quantities.   The agreement with the mass ratio
derived from the inclination is convincing.

The CCF  profile of the star B  is displaced by the  orbital motion by
less  than its full  half-width.  This  means that  some areas  on the
stellar  surface move  in anti-phase  with the  orbit. Indeed,  in the
phased CCF  profiles in Figure~\ref{fig:B}, the systemic  RV marked by
the   vertical  dotted   line  is   always  within   the   dip.   The
center-of-gravity of the system Ba,Bb is located inside the star Ba at
a distance of $ a q /(1+q) = 0.83 \; R_\odot$ from the center. Despite
the  short period,  the  close  binary system  is  well detached.  The
rotation is much slower than the break-up speed of $\sim$400 \kms. 

 Spots  on  the rotating  star  Ba  cause  variation of  its  apparent
 RV. However, the semi-amplitude of the orbital motion, 54.93 \kms, is
 substantially  larger  than  any   plausible  effect  of  the  spots.
 Moreover,      the      variable      spot     distribution      (see
 Figure~\ref{fig:lightcurve}) would produce  an irregular and variable
 RV curve, while the actual RV variation is sinusoidal and coherent on
 a time span of several years.

\begin{figure}
\plotone{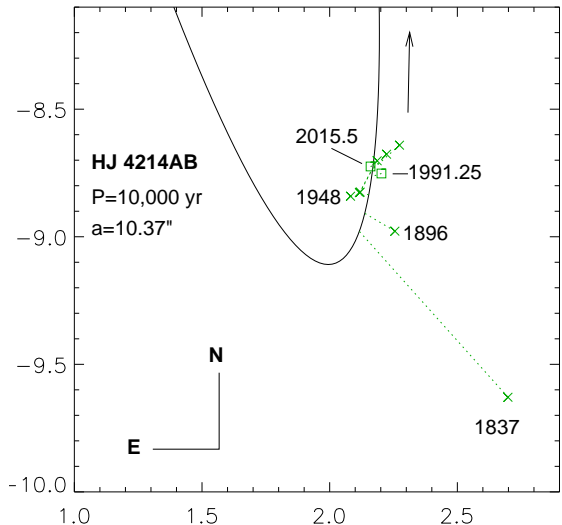}
\caption{Selected  positions of  the outer  pair A,B  in the  sky. The
  dates are  marked; squares denote  data of {\it Hipparcos}  and {\it
    Gaia}, crosses  are other measurements.   Dotted lines connect
    measurements to the expected positions on the orbit.  The line is
  a  fragment  of the  tentative  orbit.   Axis  scale in  arcseconds,
  component A is at the coordinate origin, outside the plot.
\label{fig:AB}
}
\end{figure}

\section{Outer Orbit}
\label{sec:AB}

The projected separation  between A and B is $s=614$  au. If the orbit
is circular  and oriented face-on,  $s$ equals the semimajor  axis and
the third Kepler  law gives a period $P^* \approx 8000$  yr for a mass
sum  of 3.5 \msun.   The orbital  speed is  then $2  \pi s/P^*  = 0.5$
au~yr$^{-1}$ or 2.4 \kms ~or  7.3 mas~yr$^{-1}$. If the pair A,B moved
with this  speed, its position would have  changed by 1\farcs3  in 180 years
elapsed since  its discovery  in 1837; such a  displacement is
measurable.  In  fact  the pair  moved much  less, suggesting
that the  actual period  might be longer  and/or the  apparent orbital
motion   in  the   plane  of   the  sky   is  reduced   by  projection
(Figure~\ref{fig:AB}).   The  historical  micrometer and  photographic
measurements  recorded in  the WDS  have large  random  and systematic
errors and do  not help in elucidating the orbital  motion of the wide
pair,  apart  from the  fact  that  it  is substantially  slower  than
expected for a face-on circular orbit. Modern measurements of A,B by
\citet{Vogt2012} and \citet{Tokovinin2010} also appear somewhat
discordant, likely because of imperfect calibration. 

The most accurate measurements of the relative position of the stars A
and  B (where  A  is the  photo-center  of Aa,Ab)  are available  from
Hipparcos and  {\it Gaia}  on a time  base of  24.25 yr. The  pair has
moved  by $-$37.0\,mas  and +20.4\,mas  in the  radial  and tangential
directions, respectively.  In 2015.5  the relative position of A,B was
194\fdg13  and  8\farcs988.  The  relative  positions  of  A,B can  be
described  approximately  by  a  tentative orbit  ($P=10^4$ yr,  $T=120$,
$a=10\farcs4$,  $e=0.2$, $\Omega =  11\fdg5$, $\omega  = 295\degr$,
$i=85\degr$).   However, the  orbit remains  essentially unconstrained
and  these arbitrary elements  have little  value.  According  to this
orbit,  the motion  is direct  (position angles  increase  with time),
but  the   {\it  Hipparcos}  and  {\it   Gaia}  positions,  after
correction for precession, suggest a retrograde (clock-wise) motion. 

{\it  Gaia}  measured  accurate  PMs  of  the stars  A  and  B;  their
difference  can  throw  some  light  on  the  motion  in  the  wide
pair. However, the orbital motion of  Aa,Ab with a 20.1 yr period must
be subtracted.  The  photo-center displacement of A is  related to the
relative position  of Aa,Ab  on its  orbit by the  wobble factor  $f =
-q/(1+q)  + r/(1+r)$, where  $q =  0.89$ is  the mass  ratio and  $r =
10^{-0.4 \Delta m}$  is the light ratio of Aa,Ab  which depends on the
wavelength.  In  the $V$ and $K$  bands, $r$ equals 0.59  and 0.70 and
$f$ is $-0.10$ and $-0.06$,  respectively.  According to the new orbit
of Aa,Ab, in  2015.5 the component Ab moved on the  sky relative to Aa
with the velocity  of $(-15.7, +3.2)$ mas\,yr$^{-1}$ in  the R.A.  and
declination, respectively.   The photo-center velocity scaled  by $f =
-0.10$ is $(1.6, -0.3)$  mas~yr$^{-1}$. The difference between the PMs
of A  and B  measured by {\it  Gaia} is $(1.1,  -2.3)$ mas\,yr$^{-1}$.
Therefore, the  orbital velocity of B  relative to A  was $(0.5, 2.0)$
mas~yr$^{-1}$. The tentative orbit predicts relative motion of $(-0.2,
1.8)$  mas\,yr$^{-1}$.   The motion  in  declination  agrees well  and
matches the historic trend  of decreasing angular separation between A
and B (B moves  to the North toward A). The smaller  motion in R.A. is
of opposite sign. The discrepancy could be explained if the star B had
another low-mass  companion with a period  of a few  decades.  Yet, no
such companion to B was found by high-contrast imaging \citep{SPOTS}.

The measured difference  between the RVs of B and  A is  3 \kms, of  the same
order of magnitude as the  expected orbital speed of A,B. However, the
center-of-mass RV of  B is likely biased by the  asymmetry of its wide
CCF,    hence   the    measured   RV   difference    cannot   be
trusted. Qualitatively, it  hints that the orbital motion  in the wide
pair A,B is directed mostly along the line of sight, while the motion
in the plane of the sky is slower, as observed.

\section{Photometry, Lithium, and Age}
\label{sec:model}

\begin{figure}
\plotone{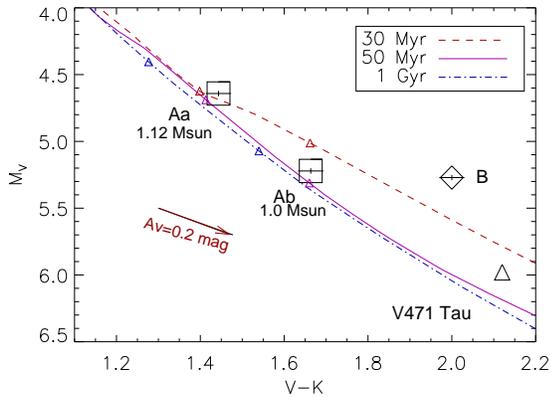}
\caption{Location  of  the  components  Aa  and Ab  (squares)  on  the
  color-magnitude   diagram.    The   lines  are   PARSEC   isochrones
  \citep{PARSEC} for solar metallicity and ages of 30 Myr, 50 Myr, and
  1 Gyr; small triangles mark the masses of 1.12 and 1.0 \msun ~on the
  isochrones.   The locations  of  B  and V471~Tau  are  shown by  the
  diamond  and triangle,  respectively. The  arrow corresponds  to the
  interstellar extinction $A_V = 0.2$ mag.
\label{fig:cmd}
}
\end{figure}

As the components A and  B are separated by 9\arcsec, their individual
photometry is readily available. The ratio  of the CCF areas of Aa and
Ab leads to the magnitude difference $\Delta V_{\rm Aa,Ab} = 0.73$ mag
after  correcting   for  the  dependence  of  the   line  contrast  on
temperature.  The  latest speckle interferometry  gives $\Delta V_{\rm
  Aa,Ab} = 0.58$ mag, which I adopt here.  The differential photometry
of Aa,Ab  by \citet{Vogt2012}  is $\Delta K_{\rm  Aa,Ab} =  0.34$ mag,
while \citet{Tokovinin2010} measured $\Delta K_{\rm Aa,Ab} = 0.38$ mag
and $\Delta  H_{\rm Aa,Ab}  = 0.36$ mag.   Therefore, the $V$  and $K$
magnitude   of  three   resolved  components   Aa,  Ab,   and   B  are
measured.  Errors of $\pm$0.02  mag in the magnitude and $\pm$0.03
  mag in the color index are assumed. 

\begin{deluxetable}{l c cc  }
\tabletypesize{\scriptsize}     
\tablecaption{Parameters of components
\label{tab:model} }  
\tablewidth{0pt}                                   
\tablehead{                                                                     
\colhead{Parameter} & 
\colhead{Aa} &
\colhead{Ab} &
\colhead{B} 
}
\startdata
$V$ (mag) & 8.81  & 9.39 & 9.44 \\
$V-K$ (mag) & 1.44 & 1.66 & 2.00 \\
${\cal M}$ (${\cal M}_\odot$) & 1.12 & 1.0 & 1.0: \\
$T_{\rm eff}$ (K) & 5900 & 5590 & 5250: 
\enddata
\end{deluxetable}

Using the {\it Gaia} distance modulus 4.169$\pm$0.005 mag and assuming
zero  extinction, the  components  are placed  on the  color-magnitude
diagram (CMD)  in Figure~\ref{fig:cmd}.  The components Aa  and Ab are
located  very  close  to  the  main sequence.   For  reference,  three
isochrones  from  \citet{PARSEC} for  solar  metallicity are  plotted.
 The Dartmouth isochrones  \citep{Dotter2008} were also tested. The 1 Gyr
isochrone is similar to the PARSEC one, while the 50~Myr Dartmouth isochrone 
does not match the data.

Table~\ref{tab:model} gives the  measured magnitudes of three resolved
components,  the measured  masses  of  Aa and  Ab,  and the  effective
temperatures deduced from the $V-K$ colors and isochrones.  {\it Gaia}
gives effective  temperatures of 5760\,K  for A (close to  the average
temperature  derived  here)  and  5242\,K for  B.   The  corresponding
spectral  types  are  G0V,  G6V,  and  K0V    \citep{Pecaut2013}.
Curiously, {\it Gaia} found an extinction  of 0.45 and 0.32 mag in the
$G$ band, assuming that A and  B are single stars.  However, there are
no obvious  interstellar adsorptions in  the spectrum of A.  Given
the distance of 68\,pc, an average extinction of $A_V = 0.1$ mag can be
expected. 

The colors,  absolute magnitudes,  and masses of  Aa and Ab  match the
standard values of  the main-sequence stars of spectral  types G0V and
G6V, within the uncertainty of  both the data and the isochrones.  The
areas of  the CCF dips  of Aa and  Ab correspond approximately  to the
stars of  solar metallicity.  Small  triangles in Figure~\ref{fig:cmd}
show  locations  of  stars  with  the  masses of  Aa  and  Ab  on  the
isochrones.  The 50 Myr isochrone  gives the best match, while at
  the  1 Gyr  age  the stars  should  be $\sim$0.2  mag brighter  than
  observed. However, an extinction of $A_V = 0.2$ mag would remove the
  discrepancy.  On  the other hand, the rapidly  rotating star B does
not conform to the standard isochrones.

\begin{figure}
\plotone{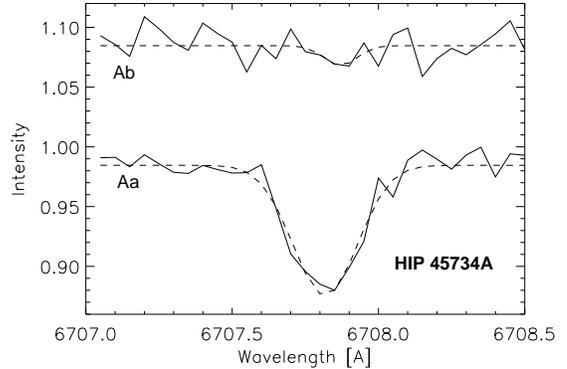}
\caption{Profiles of  the lithium lines  in the components  Ab (upper)
  and  Aa (lower) are  plotted by  full lines,  displaced vertically.
  The dotted lines are fitted Gaussians.
\label{fig:lithium}
}
\end{figure}

I do not see the 6708\AA ~lithium line in the spectrum of B because of
its  fast  rotation  and   correspondingly  low  line  contrast.   The
narrow-lined components Aa and Ab  are resolved in the CHIRON spectra.
Using  the  technique of  \citet{paper1},  the  lithium  line in  each
component  can  be measured  separately.   Quite  surprisingly, it  is
absent    in    the    component    Ab,   but    prominent    in    Aa
(Figure~\ref{fig:lithium}).   Its equivalent  width  (EW) is  30$\pm$3
m\AA, less than 60\,m\AA  ~measured by \citet{SACY} from the unresolved
spectrum of A.  The component Aa contributes $\sim$0.6 fraction of the
flux, so the intrinsic EW of its lithium line is $\sim$50\, m\AA.

At the age  of the Pleiades, $\sim$150\,Myr, stars  with the masses of
Aa  and  Ab  have  strong  lithium  lines  with  an  EW  of  120\,m\AA
~\citep{Soderblom1993}.  In the Hyades, the EW is about 80 and 34 m\AA
~for stars with $B-V$ of  0.6 and 0.7 mag \citep{Soderblom1993b}.  The
presence of  lithium in Aa and  its depletion in Ab  suggest that this
system is substantially  older than the Pleiades and  maybe even older
than the  Hyades. However, lithium  depletion depends not only  on the
age;  this  uncertainty  prevents  age-dating of  HIP~45734  based  on
lithium.   The  isochrones in  Figure~\ref{fig:cmd}  favor  an age  of
$\sim$50\,Myr, but  their dependence  on metallicity and  the existing
uncertainty  of  the stellar  evolutionary  models  preclude any  firm
conclusions.   An interstellar extinction  of 0.2 mag  would place
  the stars  Aa and  Ab on  the 1 Gyr  isochrone.   Moderate projected
axial  rotation  of the  components  Aa  and  Ab, $\sim$4  \kms,  also
suggests that these stars are not very young.

The component B (i.e the combined light of Ba and Bb) is located above
the main sequence.  If the  small contribution of the star Bb, assumed
to be an M3V dwarf, is accounted  for, Ba moves in the CMD to the left
 by  $\sim$0.15 mag, but  still remains above the  main sequence.
The characteristics  of Ba,Bb are reminiscent of  post common envelope
binaries,  where a  main-sequence dwarf  is  paired to  a white  dwarf
(WD). The  prototypical example, V471~Tau in the  Hyades, has spectral
type  K1V, orbital period  0\fd52, and  a DA-type  secondary component
\citep{Vaccaro2015}, readily detectable in  the UV.  The above authors
established that the  mass of the red dwarf in V471~Tau  is close to 1
\msun,  contradicting its  late spectral  type and  the  low effective
temperature of 5020\,K.  Obviously,  this post common envelope star is
not a normal  dwarf.  The same may be  true regarding HIP~45734B.  The
position of V471~Tau in the CMD is marked by the large triangle.

\section{Discussion}
\label{sec:sum}

The {\it Gaia} distance and  proper motion, together with the systemic
radial  velocity  of  A,  4.8  \kms, correspond  to  the  heliocentric
Galactic velocity  vector of $(U,V,W)  = (-36.1, -15.9,  -12.5)$ \kms.
This velocity does not match  any known kinematic group of young stars
in the solar  neighborhood.  For example, the Hyades  moving group has
$(U,V,W) = (-39, -18, -2)$  \kms.  The object certainly belongs to the
Galactic disk.  \citet{Tetzlaff2011}  proposed that HIP~45734 could be
ejected  from the Tucana-Horlogium  association.  However,  the weakly
bound A,B pair  would survive only if the  ejection velocity were less
than  its orbital  velocity  (2 \kms),  making  the ejection  scenario
unlikely.

I looked for potential common  proper motion companions in {\it Gaia}.
All sources  within 3\degr ~radius  with parallax larger  than 10\,mas
were retrieved;  none matches the  parallax and PM of  HIP~45734.  The
best  candidate (09:42:33.853,  $-$76:23:54.03),  at 1\fdg84  distance
from HIP~45734,  has a PM of $(-116.9,  65.9)$ mas~yr$^{-1}$, parallax
11.3\,mas, and $G=18.08$ mag, hence it is not associated.

The secondary component in the Ba,Bb system is not detected spectrally
or photometrically.  It could be  either a low-mass dwarf  of spectral
type  M3V or  a white  dwarf. In  the latter  case, the  short orbital
period of Ba,Bb is a result  of the common envelope evolution, like in
the prototypical binary  V471~Tau \citep{Vaccaro2015}.  Another recent
example of such  object is Wolf~1130 (HIP~98906), where  a late-M type
subdwarf is  paired with an  unusually massive WD  on a 0.5  day orbit
\citep{Mace2018}.    Old    post   common   envelope    binaries   are
chromospherically active and often  masquerade as young stars.  If the
component Bb  is a WD, it  could be detected  in the UV.  In  the {\it
  Galex} image  of HIP~45734  (wavelength 0.231 $\mu$m),  the southern
star B is brighter than  the star A. However, chromospherically active
stars are also  strong UV and x-ray emitters. In  short, the nature of
the star Bb remains unknown.

HIP~45734 is  an excellent illustration  of errors that could  be made
if the  multiplicity were ignored  or overlooked. A close  binary 
Ba,Bb  has a  fast  rotation and  an  active chromosphere  and can  be
mistaken for a young single star.  An unrecognized close unresolved pair 
Aa,Ab  falsifies determination  of stellar  parameters  using standard
relations  (e.g. in  the TESS  input  catalog), while  its {\it  Gaia}
parallax is  possibly biased.   A wide binary   A,B  with a relative
motion accurately measured  by {\it Gaia} lends itself  to a dynamical
analysis of  its orbit, but in  fact its motion is  strongly biased by
the subsystem and  the estimated components' masses are  also wrong if
they are deemed to be simple stars. 

This multiple  system is less interesting than  originally thought. It
is certainly not a PMS object and is possibly older than the Hyades,
although still younger than a typical age of the Galactic disk.

\acknowledgements

I  thank   the  operators  of   the  1.5-m  telescope   for  executing
observations  of  this  program  and  the  SMARTS  team  at  Yale  for
scheduling and pipeline processing.   Re-opening of CHIRON in 2017 was
largely due  to the  enthusiasm and energy  of T.~Henry.   Two spectra
were taken on my request  by F.~Walter.  Comments by the anonymous
  Referee helped to improve the presentation.

This work  used the  SIMBAD service operated  by Centre  des Donn\'ees
Stellaires  (Strasbourg, France),  bibliographic  references from  the
Astrophysics Data  System maintained  by SAO/NASA, and  the Washington
Double  Star Catalog  maintained at  USNO.  This  paper  includes data
collected   by  the  TESS   mission  funded   by  the   NASA  Explorer
Program. This work has made use of data from the European Space Agency
(ESA) mission Gaia (https://www.cosmos.esa.int/gaia), processed by the
Gaia    Data    Processing     and    Analysis    Consortium    (DPAC,
https://www.cosmos.esa.int/web/gaia/dpac/ consortium). Funding for the
DPAC  has been provided  by national  institutions, in  particular the
institutions participating in the Gaia Multilateral Agreement.

{\it Facilities:} {CTIO:1.5m, SOAR, Gaia, TESS}


\end{document}